\documentclass[10pt, prb, aps, amssymb, twocolumn, superscriptaddress, notitlepage, longbibliography, floatfix]{revtex4-2}

\usepackage{amsmath}
\usepackage{graphicx}
\usepackage{color}
\usepackage{bbm}
\usepackage[normalem]{ulem}
\usepackage[colorlinks]{hyperref}
\usepackage{dsfont}
\usepackage{physics}
\usepackage{enumitem}

\usepackage[english]{babel}

\newcommand{\xim}{\xi_m}

\begin{document}

\title{Machine learning prediction of the convergence criterion for a topological invariant of finite non-Hermitian chains} 

\author{Raghav Chaturvedi}
\email{raghav.chaturvedi@uni-wuerzburg.de}
\affiliation{Institute for Theoretical Physics and Astrophysics, and W\"{u}rzburg-Dresden Cluster of Excellence on Complexity, Topology and Dynamics in Quantum Matter ctd.qmat, Julius-Maximilians-Universit\"{a}t W\"{u}rzburg, Am Hubland, D-97074 W\"{u}rzburg, Germany}
\date{\today}

\author{Viktor K\"{o}nye}
\affiliation{Institute for Theoretical Physics Amsterdam, University of Amsterdam,
Science Park 904, 1098 XH Amsterdam, The Netherlands}

\author{Ewelina M. Hankiewicz}
\affiliation{Institute for Theoretical Physics and Astrophysics, and W\"{u}rzburg-Dresden Cluster of Excellence on Complexity, Topology and Dynamics in Quantum Matter ctd.qmat, Julius-Maximilians-Universit\"{a}t W\"{u}rzburg, Am Hubland, D-97074 W\"{u}rzburg, Germany}

\begin{abstract}

A topological invariant based on polar-decomposition of matrices correctly captures the topology of finite non-Hermitian chains exhibiting the non-Hermitian skin effect, provided that an appropriate crop-length parameter is chosen. This parameter, which sets the cutoff used in the calculation of the invariant, is usually chosen empirically and becomes especially important near topological phase transitions, where finite-size effects are strongest. Here we show that the required crop-length is controlled by physical decay (localization) lengths. For nearest-neighbor and pure longer-range hopping Hatano-Nelson-type chains, the crop-length is set mainly by a single localization length and is well approximated by a scalar multiple of that length. For more general longer-range hopping models, it is governed instead by a multichannel root structure of the characteristic polynomial. Random-forest regression captures finite-size and near-boundary corrections while preserving this decay-length interpretation. Trained on one set of Hamiltonians, the predictor accurately generalizes to unseen Hamiltonians and complex base energies, reproducing crop-lengths across full phase diagrams. We further show that the predictions learned from clean nearest-neighbor hopping chains remain stable under moderate hopping disorder. These results provide a practical and physically interpretable way to choose the crop-length, which in turn determines when the real-space invariant can reliably capture the topology of finite non-Hermitian chains.  

\end{abstract}

\maketitle

\section{Introduction} 

\label{sec:intro}

Quantum mechanics restricts the Hamiltonian of a closed system to be Hermitian ($H = H^\dagger$) with real eigenvalues and orthogonal eigenvectors \cite{sakurai1994}. Systems coupled to an external environment can be dissipative and are not restricted to be Hermitian. Such open quantum systems can be described microscopically, for instance using the Lindblad dynamics \cite{sudarshan1976, Lindblad1976-cd}, but an alternative effective description is to represent them, whenever possible, using effective non-Hermitian Hamiltonians \cite{Ashida2020, Bergholtz2021, Gong2018, Okuma2023}.
The eigenvalue spectra of non-Hermitian Hamiltonians are generally complex, which changes the notion of topology compared to the Hermitian case. The topology of non-Hermitian Hamiltonians has been extensively studied, leading to a 38-fold symmetry classification \cite{Kawabata2019}. 

In practice, the dissipation (gain or loss) in these systems can be fine-tuned to realize topological non-Hermitian phases. Several platforms where this has been realized include atomic systems \cite{Xu2017, Lee2014, Li2019, Yamamoto2019, Nakagawa2018}, electronic circuits \cite{Hofmann2020, Helbig2020}, optical and photonic systems \cite{Makris2008, Schomerus2013, Zhen2015, Malzard2018, Cerjan2019, Chen2017}, mechanical, acoustic, and other metamaterials  \cite{Liu2021, Weidemann2020, Brandenbourger2019, Ghatak2020, Zhang2021a, Zhang2021b}, 
as well as non-Hermitian conductance responses of quantum Hall and quantum spin Hall devices \cite{ochkan2025, chaturvedi2025, ozer2024, chaturvedi2026qsh}. The responses associated with non-Hermitian topological phases have several potential applications, for instance, in sensing \cite{Chen2017, Hodaei2017-tx, Budich2020, könye2023nonhermitian}, directional amplification \cite{Wanjura2020-fr, Ramos2021}, and light funneling \cite{Weidemann2020}.

A central feature of non-Hermitian topology is the existence of point gaps: the eigenvalue spectrum of a non-Hermitian Bloch Hamiltonian $H(k)$ in the complex-energy plane avoids a chosen base energy $E_B$~\cite{Gong2018,Kawabata2019,Bergholtz2021}. 
In one dimension, this allows the complex spectrum to wind around $E_B$. 
The corresponding winding number is
\begin{equation}
w(E_B)=
\frac{1}{2\pi i}
\int_0^{2\pi} dk\,
\partial_k \log \det[H(k)-E_B].
\label{eq:kspace_winding_general}
\end{equation}
For a one-band model, the determinant is unnecessary and $w(E_B)$ simply counts the number of times the complex energy curve winds around $E_B$~\cite{Gong2018,Bergholtz2021}. 
A minimal example is the Hatano-Nelson (HN) model: a one-band chain with non-reciprocal nearest-neighbor hopping~\cite{Hatano1996,Hatano1997,Hatano1998}. In this model, the periodic-boundary spectrum forms a loop in the complex-energy plane, and nonzero winding is associated with the exponential localization of an extensive number of open-chain eigenstates toward one boundary, namely the non-Hermitian skin effect~\cite{Yao2018,Okuma2020,Bergholtz2021,Hughes2021}. 

When a momentum-space description is not possible, for instance in finite open systems or in disordered systems without translational invariance, real-space invariants are needed to characterize the topology. One such quantity is the polar-decomposition real-space topological invariant $w_{\rm PD}$ \cite{Hughes2021, Ochkan2024,chaturvedi2025,ozer2024,chaturvedi2026qsh}, which approaches a non-zero integer in nontrivial point-gapped phases. 

This invariant has been used successfully to characterize the topology of finite one-dimensional non-Hermitian HN chains with open boundary conditions and finite-size non-Hermitian conductance matrices. The real-space invariant reproduces the clean momentum-space winding number in the translationally invariant limit. 
For a finite one-dimensional system of length $N$ with real-space Hamiltonian matrix $H$, one writes the polar decomposition
\begin{equation}
H - E_B = QP,
\label{eq:polar_decomposition}
\end{equation}
where $P=[(H-E_B)^\dagger (H-E_B)]^{1/2}$ is positive semidefinite and $Q$ is the unitary part of the polar decomposition. For numerical calculations, we define the local marker  
\begin{equation} \label{eq:diagonal}
d_n =
\left[Q^\dagger [Q,X]\right]_{nn}
=
\left[X-Q^\dagger XQ\right]_{nn},
\end{equation}
where $X$ is the position operator (a diagonal matrix with entries $[0, 1, ..., N-1]$). Using the crop-length parameter $\ell$, we average $d_n$ over the central window to get the real-space invariant ~\cite{Hughes2021} 
\begin{equation}
w_{\rm PD}(\ell)
=
\frac{1}{N-2\ell}
\sum_{n=\ell+1}^{N-\ell} d_n ,
\qquad 0\leq \ell < N/2 .
\end{equation}
The crop-length parameter fixes the bulk window used in the trace and reduces edge-dependent contributions.

The required value of $\ell$ is usually chosen
  empirically, and this choice becomes more important near a topological
  transition. We define 
  \begin{equation}
\Delta(\ell)
=
\left|w_{\rm PD}(\ell)-w\right|,
\end{equation}
where $w$ is computed independently from the clean momentum-space Hamiltonian
  using Eq.~\eqref{eq:kspace_winding_general}. For a chosen tolerance
  $\epsilon$, the minimum crop-length is
  \begin{equation}
  \ell_\star(\epsilon)
  =
  \min\{\ell:\Delta(\ell)<\epsilon\}.
  \end{equation}
  We treat the prediction of $\ell_\star$ as a supervised machine-learning
  problem. Here a sample means one finite-chain instance, specified by its system size,
  hopping parameters, and base energy when applicable. The input is a set of parameters (features) for a finite-sized chain, and the output is either a yes/no prediction of whether the tolerance condition $\Delta(\ell)<\epsilon$ can be reached within the allowed crop range, or the numerical value of the required crop-length.

The paper is organized as follows. Section~\ref{sec:nn_hn} treats the
  nearest-neighbor Hatano-Nelson chain at $E_B=0$. Section~\ref{sec:mhop_hn} extends the same analysis to pure $m$-hop non-Hermitian chains. In Section~\ref{sec:mixed_model} we consider a non-Hermitian chain with both first- and second-neighbor hoppings. Next, in Section~\ref{sec:complex_EB_extension} we extend the analysis to complex energy base points $E_B$. We also test the same on finite-range
  clean hopping models with up to third-, fourth-, and fifth-neighbor hoppings. Finally in Section~\ref{sec:disorder_robustness} we test whether the clean
  nearest-neighbor crop predictor remains useful under hopping disorder. All machine-learning calculations are implemented using the Python package
  scikit-learn~\cite{pedregosa2018}.

\section{Hatano-Nelson model}
\label{sec:nn_hn}

\begin{figure}[t]
\centering
\includegraphics[width=0.95\linewidth]{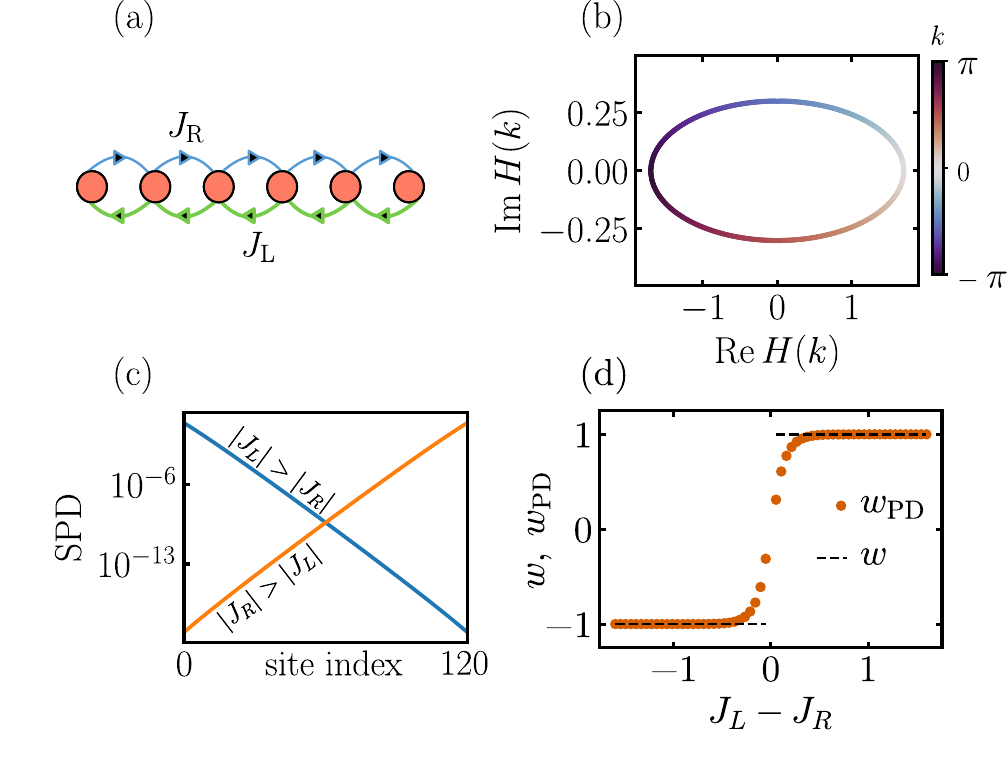}
\caption{(a) Finite-size HN chain with asymmetric nearest-neighbor hoppings $J_R$ and $J_L$. (b) Spectrum of the Bloch Hamiltonian $H(k)$ [Eq.~\eqref{eq:hn_nn_bloch}] in the complex-energy plane. The color scale follows the momentum ($k$) across the Brillouin zone, tracing an ellipse. Here $(J_R, J_L) = (1,0.6)$. (c) Summed probability density (SPD) of the open chain plotted on a log-linear scale against the site index. The non-Hermitian skin effect occurs at opposite boundaries for $|J_L|>|J_R|$ and $|J_R|>|J_L|$. (d) Real-space polar-decomposition invariant $w_{\rm PD}$ across the topological phase transition compared with the exact clean winding number $w$. Here $\ell=10$ and the chain length is $N=120$. 
}
\label{fig:hn_basic}
\end{figure}

We begin with the simplest one-dimensional model with nontrivial point-gap topology: the Hatano–Nelson model~\cite{Hatano1996,Hatano1997,Hatano1998,Gong2018,Bergholtz2021}. In real space, the model is described as 
\begin{equation}
H
=
\sum_{n}
\left(
J_R c_{n+1}^{\dagger}c_n
+
J_L c_n^\dagger c_{n+1}
\right),
\label{eq:hn_nn_real_space}
\end{equation}
where $c_n^\dagger$ and $c_n$ are creation and annihilation operators on lattice site $n$, and $J_R$ and $J_L$ denote the asymmetric hopping amplitudes towards the right and left, respectively (see Fig.~\ref{fig:hn_basic}(a) for a schematic of finite-size HN chain under open boundary conditions). For an infinite translationally invariant chain the Hamiltonian in momentum space reads
\begin{equation}
H(k)=J_R e^{-ik}+J_L e^{ik}.
\label{eq:hn_nn_bloch}
\end{equation} 
The energy eigenvalues of this model describe an ellipse (closed loop) in the complex-energy plane and wind around the origin ($E_B = 0$) in a clockwise (anticlockwise) sense for $|J_R| > |J_L|$ ($|J_L| > |J_R|$) [see Fig.~\ref{fig:hn_basic}(b)]. With the convention in Eq.~\eqref{eq:hn_nn_bloch}, the clean momentum-space winding is
\begin{equation}
w=
\begin{cases}
+1, & |J_L|>|J_R|,\\
-1, & |J_R|>|J_L|.
\end{cases}
\label{eq:nn_winding}
\end{equation}         

A phase transition between the two point-gapped phases occurs at $|J_R|=|J_L|$, where the loop crosses $E_B=0$ and the point gap closes.

\begin{figure}[t]
\centering
\includegraphics[width=0.95\linewidth]{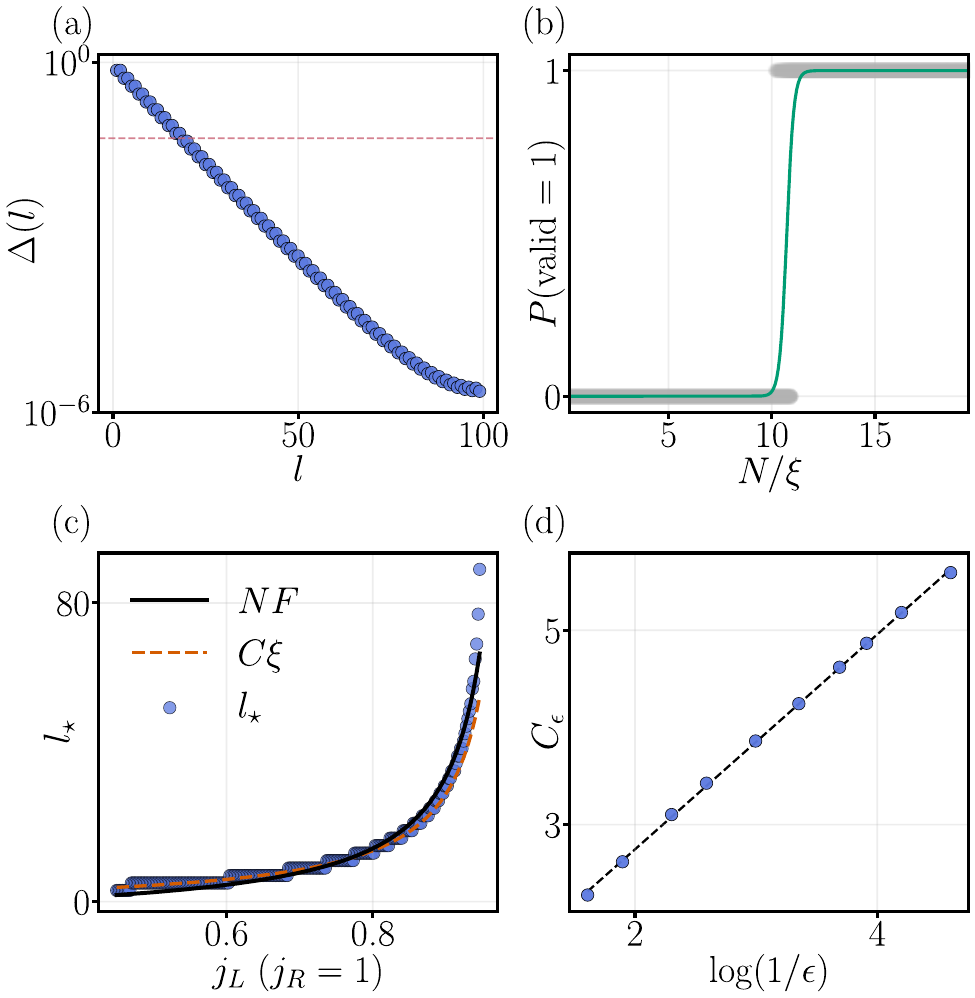}
\caption{
(a) Convergence of the real-space polar-decomposition invariant with boundary crop-length $\ell$. The plotted quantity is $\Delta(\ell)=|w_{\rm PD}(\ell)-w|$, where $w$ is the clean momentum-space winding. The horizontal dashed line marks the tolerance $\epsilon=0.05$. 
(b) Binary valid-crop labels as a function of the scaling variable $x=N/\xi$. Grey points show the raw valid/invalid labels, and the green curve shows the one-feature logistic fit $P_{\rm valid}(x)$.
(c) Fixed-size crop-length $\ell_\star$ as a function of $J_L$ for $J_R=1$ and $N = 200$. Blue points denote numerical values, the black curve shows the finite-size fit $\ell_\star=NF(x)$ with $F(x)=a/(x+b)+c$, and the orange dashed curve shows the asymptotic guide $\ell_\star=C_\epsilon\xi$.
(d) Tolerance dependence of the effective prefactor $C_\epsilon$. The approximately linear dependence on $\log(1/\epsilon)$ is consistent with an exponentially decaying finite-size correction controlled by the localization length.
}
\label{fig:nn_crop_scaling}
\end{figure}

For a finite-size chain of length $N$ with open boundary conditions, this hopping asymmetry produces the non-Hermitian skin effect \cite{Bergholtz2021, Gong2018}. To see this, we evaluate the summed probability density
\begin{equation}
{\rm SPD}(n)=\sum_{\alpha}|\psi_\alpha(n)|^2,
\label{eq:spd_definition}
\end{equation}
where $\psi_\alpha(n)$ are normalized right eigenvectors of the open chain Hamiltonian matrix. As shown in Fig.~\ref{fig:hn_basic}(c), the two topological phases produce exponential localization of eigenstates at opposite ends of the chain. The momentum-space winding $w$ and the real-space invariant $w_{\rm PD}$ distinguish these two phases, as shown in Fig.~\ref{fig:hn_basic}(d). Moreover, for a fixed chain length $N$, the $w_{\rm PD}$ invariant converges to $w$ as $\ell$ is increased, as shown in Fig.~\ref{fig:nn_crop_scaling}(a).

 We now consider the crop-length problem for this model as a supervised-learning task. Each instance of a finite open chain, or sample, is specified by $(N,J_R,J_L)$. For each sample, we compute $w_{\rm PD}(\ell)$ as a function of $\ell$ and compare it with the exact clean value $w$ from Eq.~\eqref{eq:nn_winding}. A sample is labeled valid if $\Delta(\ell)<\epsilon$ for at least one allowed crop-length.

  We use random-forest (RF) models~\cite{Breiman2001-be}: a classifier for the valid/invalid decision and a regressor for the prediction of $\ell_\star$ on valid samples. The input features include the raw parameters, simple non-reciprocity quantities, and a decay length, which we define below.

Away from the boundaries, a single-particle wavefunction amplitude satisfies
\begin{equation}
E\psi_n = J_R\psi_{n-1}+J_L\psi_{n+1}.
\label{eq:nn_realspace_bulk}
\end{equation}
We consider an exponential component
\begin{equation}
\psi_n \propto \beta^n ,
\label{eq:nn_beta_ansatz}
\end{equation}
where $\beta$ is the spatial multiplier between neighboring sites. Substituting this form into Eq.~\eqref{eq:nn_realspace_bulk} gives
\begin{equation}
E=J_R\beta^{-1}+J_L\beta .
\label{eq:nn_h_beta}
\end{equation}
Writing $\beta=|\beta|e^{iq}$, the ansatz becomes $\psi_n\propto |\beta|^n e^{iqn}$. Thus the phase of $\beta$ gives the oscillatory part, while $|\beta|^n$ gives the spatial envelope. If $|\beta|=1$, the envelope is constant and the component is extended. If $|\beta|\neq 1$, the envelope grows or decays exponentially with position. At the base energy $E_B=0$, Eq.~\eqref{eq:nn_h_beta} gives
\begin{equation}
J_L\beta^2+J_R=0,
\label{eq:nn_root_equation}
\end{equation}
and hence
\begin{equation}
|\beta|=\left|\frac{J_R}{J_L}\right|^{1/2}.
\label{eq:nn_beta_modulus}
\end{equation}
The corresponding density profile scales as $|\psi_n|^2\sim |\beta|^{2n}$, so the density decay length is
\begin{equation}
\xi
=
\frac{1}{2|\log|\beta||}
=
\frac{1}{|\log|J_R/J_L||}.
\label{eq:xi_nn}
\end{equation}
This motivates the dimensionless feature
\begin{equation}
x=\frac{N}{\xi}
=
N|\log|J_R/J_L|| .
\label{eq:nn_x_variable}
\end{equation}

The full engineered feature set used for the machine-learning problem is
\begin{equation}
  \begin{aligned}
  \mathcal{F}_{\rm NN}
  =
  \{&
  N,\; J_R, \; J_L,\; J_L-J_R,\; |J_L-J_R|,\; J_L/J_R,\\
  &\left|\log\left|J_R/J_L\right|\right|,\; \xi,\; x=N/\xi
  \}.
  \end{aligned}
  \label{eq:nn_feature_set}
  \end{equation}
Here engineered means that, in addition to the raw hopping parameters, we also provide physically motivated combinations such as the decay length and the
  scaling variable $x=N/\xi$.

The regressor uses the full engineered feature set $\mathcal{F}_{\rm NN}$, whereas the classifier uses only the scaling variable $x=N/\xi$. To test whether the RF models generalize, we use an $80/20$ train-test split, fitting each model on $80\%$ of the samples and evaluating it on the held-out $20\%$. Since the finite-size error shows a weak residue-class dependence on $N$, we perform this split separately within each finite-size branch (see Appendix~\ref{app:finite_size_branches}); for the nearest-neighbor chain, a branch is a fixed value of $N\bmod 2$. We report the mean and standard deviation of the performance over the two branches.

For the classification problem, we report the accuracy, defined as the fraction of test samples classified correctly:
\begin{equation}
{\rm Acc.}
=
\frac{TP+TN}{TP+TN+FP+FN},
\end{equation}
where $TP,TN,FP,FN$ are the true-positive, true-negative, false-positive, and false-negative counts on the test set. Since the valid class, labeled 1, and invalid class, labeled 0, are not equally represented, we also report the balanced accuracy,
\begin{equation}
{\rm Bal.\ Acc.}
=
\frac{1}{2}
\left[
\frac{TP}{TP+FN}
+
\frac{TN}{TN+FP}
\right].
\end{equation}
Balanced accuracy gives equal weight to the valid and invalid classes, which is useful
when one class appears more often than the other.

The classifier then predicts whether a valid crop exists with ${\rm Acc.}=0.983\pm0.009$ and ${\rm Bal.\ Acc.}=0.980\pm0.009$, where the uncertainties are the standard deviations over the finite-size branches.

Figure~\ref{fig:nn_crop_scaling}(b) shows the raw binary valid/invalid labels as a function of the localization-length scaling variable $x=N/\xi$ defined in Eq.~\eqref{eq:nn_x_variable}. 
The solid curve is a one-feature logistic fit,
\begin{equation}
P_{\rm valid}(x)
=
\frac{1}{1+\exp[-(a x+b)]},
\label{eq:nn_logistic_fit}
\end{equation}
where $P_{\rm valid}$ is the fitted probability that a valid crop exists. 
This logistic curve is not the random-forest classifier; it is a one-dimensional diagnostic showing that the transition from invalid to
  valid samples is organized primarily by $x=N/\xi$. The fitted
  transition occurs at $x\simeq 10$, indicating that, for $\epsilon=0.05$, the chain length must be roughly an order of magnitude larger than the
  localization length, $N\gtrsim 10\,\xi$, for this method to reliably recover the winding.

For the regression problem, the target, i.e. the quantity to be predicted, is
  the normalized crop-length,
\begin{equation}
y_i=\left(\frac{\ell_\star}{N}\right)_i,
\end{equation}
where $i$ labels a test sample with a valid crop.

The random-forest regressor predicts a value $\hat y_i$. We quantify the regression performance using the coefficient of determination $R^2$,
\begin{equation}
R^2
=
1-
\frac{\sum_i (y_i-\hat y_i)^2}
{\sum_i (y_i-\bar y)^2},
\label{eq:nn_r2_definition}
\end{equation}
where $\bar y$ is the mean value of $y_i$ on the test set. The $R^2$ score measures how much of the variation in the target is explained by the model, with $R^2=1$ corresponding to perfect prediction.

The regressor predicts the normalized crop-length on the test sets with $R^2=0.997\pm0.0004$, where the uncertainty is the standard deviation over the finite-size branches. Furthermore, the impurity-based feature importance of the trained regressor, which scores each input feature by how strongly it contributes to the prediction, is dominated by the finite-size scaling variable $x=N/\xi$, with importance $0.99$.

The localization length $\xi=1/|\log|J_R/J_L||$ increases as $J_L$ approaches $J_R$, that is, towards the transition. As shown in Fig.~\ref{fig:nn_crop_scaling}(c), $\ell_\star$ increases in the same regime. Since $\ell_\star$ is integer-valued, the data are step-like rather than smooth: over a finite range of $J_L$, the same integer crop-length remains optimal, and then $\ell_\star$ jumps by one or more. 

To quantify the underlying trend, we first fit a continuous localization-length estimate,
\begin{equation}
\ell_\star^{\rm cont}(\epsilon)
=
C_\epsilon \xi.
\label{eq:hn_nn_lstar_cont}
\end{equation}
The corresponding integer estimate is obtained by rounding,
\begin{equation}
\ell_\star^{\rm est}
=
{\rm round}\!\left(\ell_\star^{\rm cont}\right).
\label{eq:hn_nn_lstar_round}
\end{equation}
For the fixed-size data in Fig.~\ref{fig:nn_crop_scaling}(c), this rounded estimate reproduces the exact integer crop-length for $40.73\%$ of the points, lies within $\pm1$ site for $87.90\%$ of the points, and lies within $\pm2$ sites for $91.53\%$ of the points. Thus the direct proportionality to $\xi$ captures the main trend, but it is not the most accurate description close to the transition.

Close to the transition, the localization length becomes comparable to the system size and the simple estimate in Eq.~\eqref{eq:hn_nn_lstar_cont} does not capture the finite-size upturn. We therefore use a finite-size fit,
\begin{equation}
\frac{\ell_\star^{\rm cont}}{N}
=
F(x),
\qquad
F(x)=\frac{a}{x+b}+c,
\label{eq:nn_finite_size_fit}
\end{equation}
where $x=N/\xi$. The integer estimate is again obtained by rounding,
\begin{equation}
\ell_\star^{\rm est}
=
{\rm round}\!\left[NF(x)\right].
\label{eq:nn_finite_size_round}
\end{equation}

For the fixed-size data in Fig.~\ref{fig:nn_crop_scaling}(c), the fitted
  parameters are $(a,b,c)=(2.329,\;-5.194,\;7.00\times10^{-3}).$ With these parameters, the finite-size fit reproduces the exact integer crop-length for $46.77\%$ of the points, lies within $\pm1$ site for $90.73\%$ of the points, and lies within $\pm2$ sites for $98.39\%$ of the points. This shows that the finite-size fit describes the near-transition regime better than the direct proportionality $\ell_\star\propto\xi$.

The proportionality constant $C_\epsilon$ depends on the chosen tolerance $\epsilon$. Away from the transition, where finite-size effects are weaker, we fit the data to
\begin{equation}
\ell_\star(\epsilon)\simeq C_\epsilon \xi.
\label{eq:ceps_fit_definition}
\end{equation}
This gives one fitted value of $C_\epsilon$ for each tolerance. These values are plotted against $\log(1/\epsilon)$ in Fig.~\ref{fig:nn_crop_scaling}(d). 
The approximately linear dependence can be understood from an exponentially decaying boundary contribution,
\begin{equation}
\Delta(\ell)\sim A e^{-\ell/\xi},
\label{eq:nn_exp_error}
\end{equation}
where $A$ is a prefactor. 
The crop condition $\Delta(\ell_\star)<\epsilon$ gives
\begin{equation}
A e^{-\ell_\star/\xi}<\epsilon.
\end{equation}
Solving this inequality gives
\begin{equation}
\ell_\star>
\xi\log(A/\epsilon)
=
\xi\log(1/\epsilon)
+
\xi\log A.
\end{equation}
Thus, using $\ell_\star\simeq C_\epsilon\xi$, the prefactor should behave approximately as
\begin{equation}
C_\epsilon \simeq \log(1/\epsilon)+{\rm const.}
\label{eq:ceps_log_scaling}
\end{equation}
This gives a simple explanation for the approximately linear growth of $C_\epsilon$ with $\log(1/\epsilon)$ in Fig.~\ref{fig:nn_crop_scaling}(d). Since $\ell_\star$ grows linearly with $\log(1/\epsilon)$ at a slope $\xi$, one can read off the localization length directly from how the crop-length depends on the tolerance, without knowing the hopping amplitudes.

\section{Pure m-hop Hatano–Nelson chains}
\label{sec:mhop_hn}

We next consider pure higher-neighbor HN chains, where hopping occurs only between sites separated by $m$ lattice spacings. The tight-binding Hamiltonian in real space is described as
\begin{equation}
H_m
=
\sum_n
\left(
J_R c_{n+m}^{\dagger}c_n
+
J_L c_n^\dagger c_{n+m}
\right).
\label{eq:mhop_real_space}
\end{equation}

\begin{figure}[t]
\centering
\includegraphics[width=0.95\linewidth]{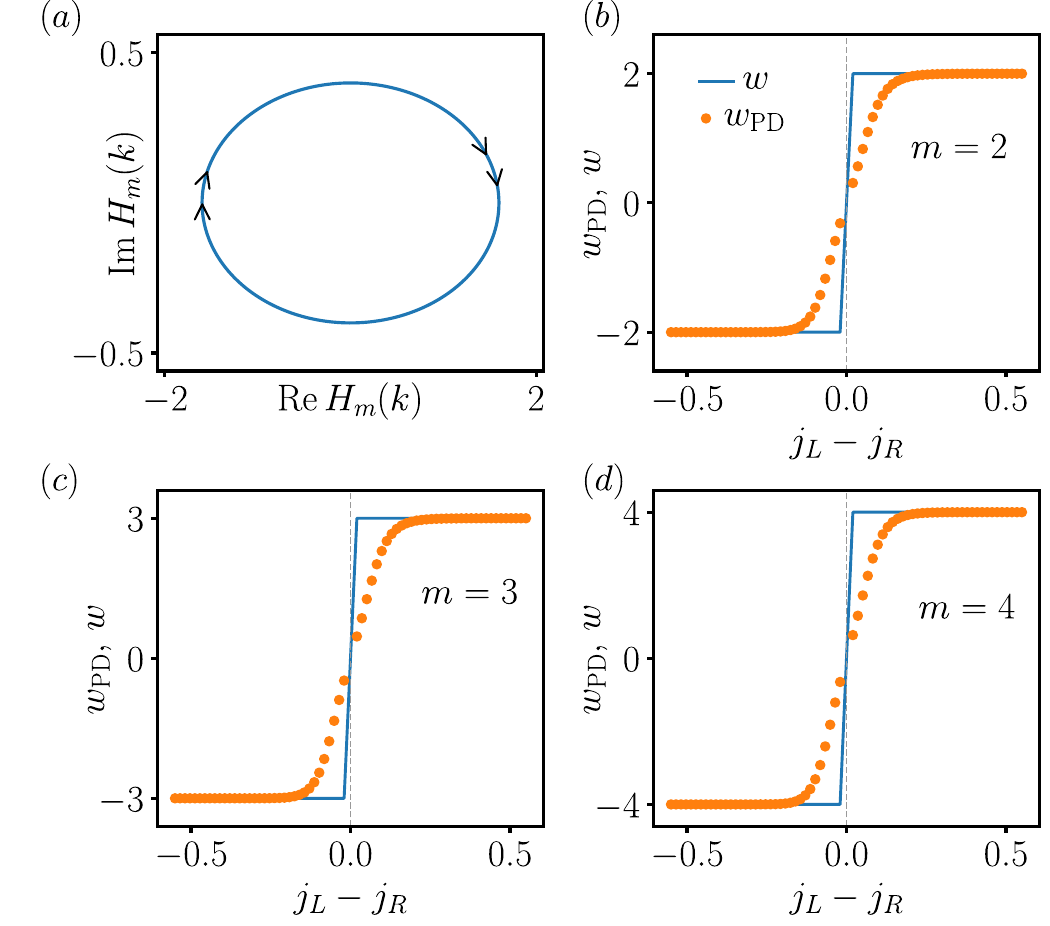}
\caption{
(a) Complex spectrum of $H_m(k)$ for $m=2$. The spectrum is an ellipse in the complex-energy plane, but it winds two times around the base point $E_B = 0$. (b-d) Real-space polar-decomposition invariant $w_{\rm PD}$ across the point-gap transition for $m=2,3,4$, compared with the exact clean winding $w$.}
\label{fig:mhop_basic}
\end{figure}

The corresponding Bloch Hamiltonian is 
\begin{equation}
H_m(k)=J_R e^{-imk}+J_L e^{imk}.
\label{eq:mhop_bloch}
\end{equation}
The spectrum is an ellipse in the complex-energy plane, but the loop winds $m$ times around the base point as $k$ goes from $0$ to $2\pi$ [see Fig.~\ref{fig:mhop_basic}(a)]. With the convention in Eq.~\eqref{eq:mhop_bloch}, the clean momentum-space topological invariant $w$ is
\begin{equation}
w=
\begin{cases}
+m, & |J_L|>|J_R|,\\
-m, & |J_R|>|J_L|.
\end{cases}
\label{eq:mhop_winding}
\end{equation}
The point-gap transition occurs at $|J_R|=|J_L|$, where the ellipse crosses $E_B=0$. The same integer winding is recovered from the real-space polar-decomposition invariant $w_{\rm PD}$ in finite open chains. This is shown in Figs.~\ref{fig:mhop_basic}(b) to \ref{fig:mhop_basic}(d) for $m=2,3,4$. 

The nearest-neighbor analysis in Sec.~\ref{sec:nn_hn} showed that the crop-length is controlled by the root-derived localization length. We now derive the corresponding localization length for the pure $m$-hop model. Replacing $e^{ik}$ by $\beta$, Eq.~\eqref{eq:mhop_bloch} becomes
\begin{equation}
H_m(\beta)=J_R\beta^{-m}+J_L\beta^m.
\end{equation}
The zero-energy condition gives
\begin{equation}
J_L\beta^{2m}+J_R=0,
\end{equation}
or
\begin{equation}
\beta^{2m}=-\frac{J_R}{J_L}.
\label{eq:mhop_roots}
\end{equation}

Thus there are $2m$ roots, but all have the same modulus,
\begin{equation}
|\beta|=\left|\frac{J_R}{J_L}\right|^{1/(2m)}.
\end{equation}
A mode $\psi_n\sim\beta^n$ has density profile $|\psi_n|^2\sim|\beta|^{2n}$, giving the density localization length
\begin{equation}
\xi_m
=
\frac{1}{2|\log|\beta||}
=
\frac{m}{|\log|J_R/J_L||}.
\label{eq:xi_m}
\end{equation}
Compared with the nearest-neighbor case ($m=1$), the localization length is therefore enhanced by a factor of $m$. For $m>1$, the hopping range is longer, but the model still has only one localization length scale. This makes it a useful test case for asking whether the crop-length scaling found for the $m = 1$ case persists for higher hopping range $m$.

The relevant engineered feature is
\begin{equation}
x_m=\frac{N}{\xim}=\frac{N|\log(J_R/J_L)|}{m}.
\label{eq:xm_definition}
\end{equation}

The ML setup is the same as in the nearest-neighbor case in Sec.~\ref{sec:nn_hn}. For each sample specified by $(m,N,J_R,J_L)$, we compute $w_{\rm PD}(\ell)$ as a function of the crop-length $\ell$ and compare it with the exact clean winding $w=\pm m$. A sample is labeled valid (1) if the error $\Delta(\ell)$ becomes smaller than the chosen tolerance $\epsilon$ before the central region is exhausted, and invalid (0) otherwise. For the pure $m$-hop models, the full feature set is
\begin{equation}
\begin{aligned}
\mathcal{F}_{m}
=
\{&
m,\; N,\; J_R,\; J_L,\; J_L-J_R,\; |J_L-J_R|,\; J_L/J_R,\\
& \left|\log\left|J_R/J_L\right|\right|,\; \xi_m,\; x_m=N/\xi_m,\; N/m
\}.
\end{aligned}
\label{eq:mhop_feature_set}
\end{equation}

As before, we avoid mixing finite-size residue branches in the train-test split. For the pure $m$-th-neighbor chain the branch structure is organized by $N\bmod 2m$ (see Appendix~\ref{app:finite_size_branches}). Within each branch, we use an $80/20$ train-test split and compute the performance separately. The reported values are the mean and standard deviation over branches.

\begin{figure}[t]
\centering
\includegraphics[width=0.95\linewidth]{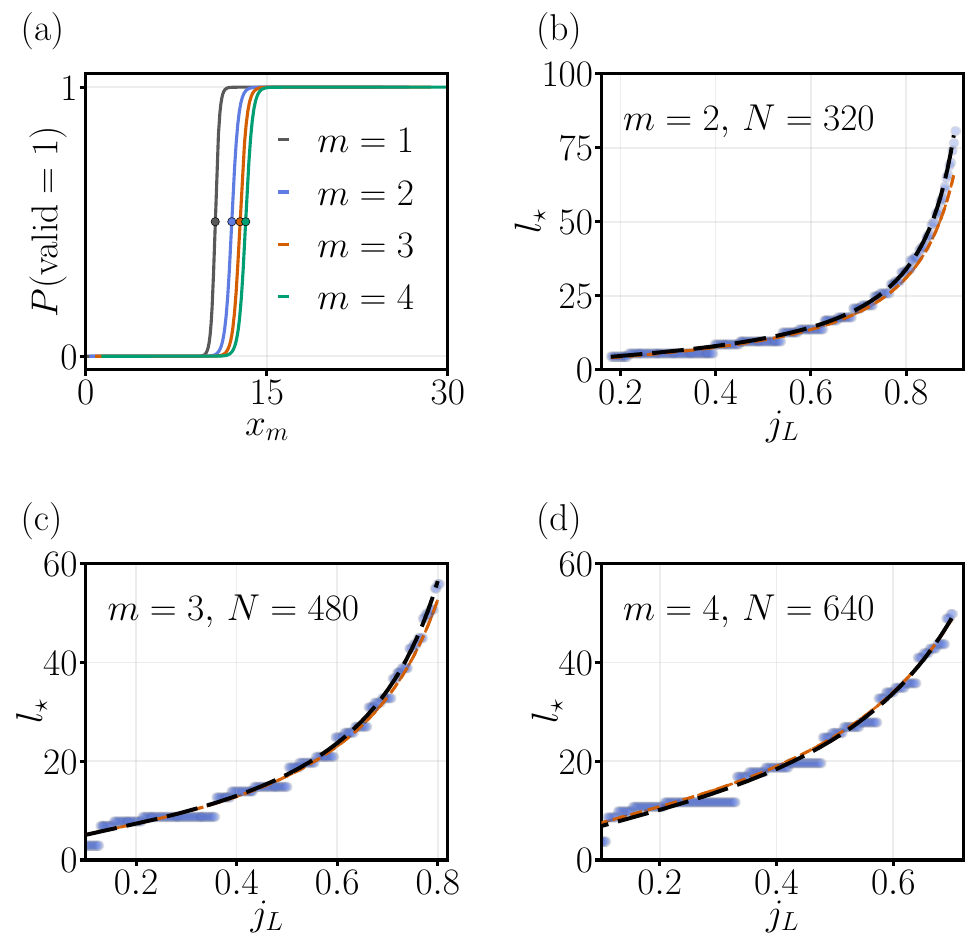}
\caption{
(a) One-feature logistic fits for the valid-crop classification problem for $m=1,2,3,4$, plotted against $x_m=N/\xi_m$. Filled circles indicate the fitted $P_{\rm valid}=1/2$ threshold points. (b-d) Integer crop-length $\ell_\star$ as a function of $J_L$ for $m=2,3,4$, with $J_R=1$. The displayed cuts use $(m,N)=(2,320),(3,480),(4,640)$. Blue points are the numerical values of $\ell_\star$. The orange dashed curves show $\ell_\star^{\rm cont}=C_\epsilon\xi_m$. The black curves show $\ell_\star^{\rm cont}=N F(x_m)$, with $F(x_m)=a/(x_m+b)+c$. Integer predictions are obtained by rounding the continuous curves.
}
\label{fig:mhop_crop_scaling}
\end{figure}

Figure~\ref{fig:mhop_crop_scaling}(a) shows one-feature logistic fits for the binary valid-crop labels for $m=1,2,3,4$ as a function of $x_m$ 
\begin{equation}
P_{\rm valid}(x_m)
=
\frac{1}{1+\exp[-(a_m x_m+b_m)]}.
\label{eq:mhop_logistic_fit}
\end{equation}
Here $P_{\rm valid}$ is the fitted probability that a valid crop exists. The filled circles in Fig.~\ref{fig:mhop_crop_scaling}(a) indicate the fitted $P_{\rm valid}=1/2$ threshold for each $m$. 

For regression, we train a random-forest regressor on the pure $m$-th-neighbor dataset at $\epsilon=0.05$. The target is again the normalized crop-length $\ell_\star/N$ for samples labeled as valid. Using the full engineered feature set and the $N\bmod 2m$ branch-aware split, the random-forest regressor predicts the normalized crop-length with $R^2=0.997\pm0.002$, where the uncertainty is the standard deviation over the finite-size branches. In terms of the integer crop-length, after rounding the predicted crop-length to the nearest integer, the prediction is exact for $90.09\pm3.50\%$ of the samples, within $\pm1$ site for $98.16\pm0.94\%$ of the samples, and within $\pm2$ sites for $99.33\pm0.44\%$ of the samples. Moreover, the feature-importance analysis gives the largest importance to $x_m=N/\xi_m$, with value $0.94$. The hopping range $m$ is the next largest feature, while all remaining features have much smaller importances.

The numerically computed values of $\ell_\star$ are shown in Figs.~\ref{fig:mhop_crop_scaling}(b)-\ref{fig:mhop_crop_scaling}(d). These panels show fixed-size cuts for $m=2,3,4$, with system sizes chosen so that $N/m$ is fixed. As in the nearest-neighbor model, $\ell_\star$ is integer-valued, and the numerical data are step-like. 

Away from the transition, the continuous localization-length estimate is
\begin{equation} \label{eq:mhop_lstar_xi}
\ell_\star^{\rm cont}(\epsilon)
=
C_\epsilon \xi_m .
\end{equation} 
Closer to the transition, we use the same finite-size fit $\ell_\star^{\rm cont}/N=F(x_m)$ as in the nearest-neighbor case [Eq.~\eqref{eq:nn_finite_size_fit}], now with $x_m=N/\xi_m$, and round to the nearest integer as before. Both these fits are shown in Figs.~\ref{fig:mhop_crop_scaling}(b)-\ref{fig:mhop_crop_scaling}(d), where the finite-size form reproduces the numerical $\ell_\star$ well across the transition. Thus the crop-length picture from the nearest-neighbor chain carries over unchanged to pure $m$-hop chains, with $\xi$ replaced by the single length $\xi_m$; this changes once several hopping ranges are present, which we consider next.

\section{Mixed $1{\rm NN}+2{\rm NN}$ Hatano-Nelson chains}
\label{sec:mixed_model}

\begin{figure}[t]
\centering
\includegraphics[width=0.95\linewidth]{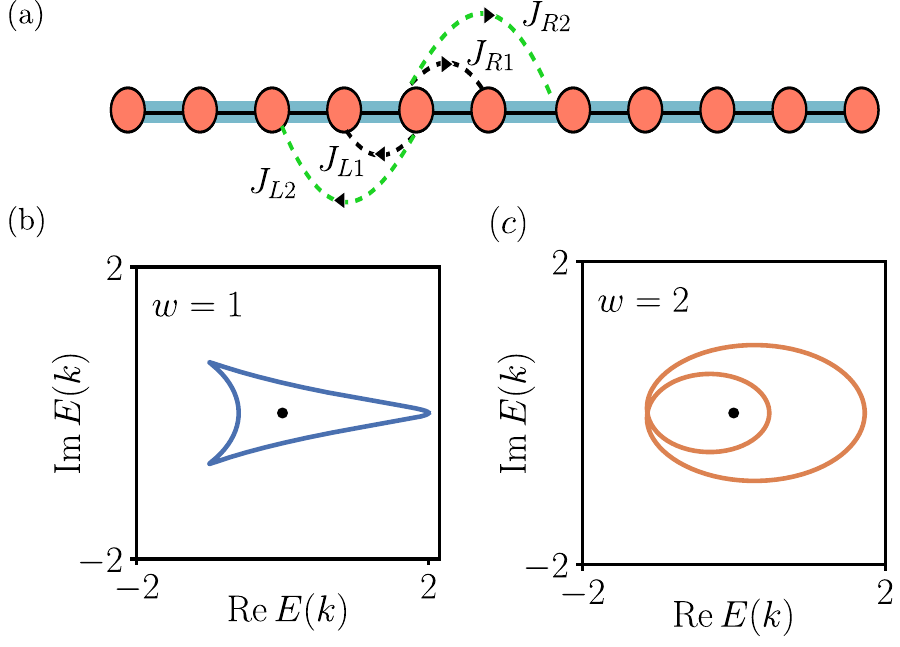}
\caption{
(a) Schematic of the open chain with first- and second-neighbor non-reciprocal hoppings.
(b-c) Point-gapped spectral loops of the Bloch Hamiltonian in the complex-energy plane. The black point marks the base energy $E_B=0$. The winding number $w$ is computed from the phase winding of $H(k)$ (see Eq.~\eqref{eq:mixed_bloch} and Eq.~\eqref{eq:kspace_winding_general}) around this base point.
}
\label{fig:mixed_complex_loops}
\end{figure}

We now consider Hatano-Nelson chains with both first- and second-nearest-neighbor non-reciprocal hoppings. The real-space Hamiltonian for an open chain is
\begin{equation}
  \begin{aligned}
  H
  =
  \sum_n \Big(
  &J_{R1}c_{n+1}^{\dagger}c_n
  + J_{L1}c_n^\dagger c_{n+1} \\
  &+ J_{R2}c_{n+2}^{\dagger}c_n
  + J_{L2}c_n^\dagger c_{n+2}
  \Big),
  \end{aligned}
  \label{eq:mixed_real_space}
  \end{equation}
where $J_{R1}$ and $J_{L1}$ are the first-neighbor hopping amplitudes to the right and left, while $J_{R2}$ and $J_{L2}$ are the corresponding second-neighbor hopping amplitudes. A schematic of an open chain is shown in Fig.~\ref{fig:mixed_complex_loops}(a). 

For an infinite translationally invariant chain, the Bloch Hamiltonian is \begin{equation}
H(k)=
J_{R1}e^{-ik}
+J_{L1}e^{ik}
+J_{R2}e^{-2ik}
+J_{L2}e^{2ik}.
\label{eq:mixed_bloch}
\end{equation} 

The spectrum of Eq.~\eqref{eq:mixed_bloch} is no longer restricted to a simple ellipse. Depending on the four hopping amplitudes, the complex-energy loop can have different non-elliptic shapes, as shown in  Fig.~\ref{fig:mixed_complex_loops}(b,c). The winding number $w$ is still the topological invariant here and the point-gap winding around $E_B=0$ is computed using Eq.~\eqref{eq:kspace_winding_general}. However, unlike the nearest-neighbor and pure $m$-th-neighbor hopping models in Secs.~\ref{sec:nn_hn} and \ref{sec:mhop_hn}, there is no single right-left hopping comparison that labels all winding sectors.

For each sample $(N,J_{R1},J_{L1},J_{R2},J_{L2}),$ we compute $w_{\rm PD}(\ell)$ and compare it with the exact clean winding $w$. For the nontrivial sectors considered below, $w\in\{-2,-1,1,2\}$. The $w=0$ sector is not included in the regression analysis because we focus on nonzero winding sectors with a nontrivial crop-length problem. A sample is labeled valid if $\Delta(\ell)$ falls below the tolerance $\epsilon$ for at least one allowed crop-length before the central region is exhausted, and $\ell_\star(\epsilon)$ is the smallest such $\ell$. 

The raw input features for this model are
$(N,J_{R1},J_{L1},J_{R2},J_{L2})$. To construct root-derived features, we replace $e^{ik}$ by $\beta$ in Eq.~\eqref{eq:mixed_bloch}, giving
\begin{equation}
H(\beta)
=
J_{R1}\beta^{-1}
+J_{L1}\beta
+J_{R2}\beta^{-2}
+J_{L2}\beta^2 .
\label{eq:mixed_beta}
\end{equation}
Multiplying by $\beta^2$ and solving at $E = E_B = 0$, we get  
\begin{equation}
J_{L2}\beta^4+J_{L1}\beta^3+J_{R1}\beta+J_{R2}=0 .
\label{eq:mixed_root_polynomial}
\end{equation}
This quartic equation has four roots, denoted $\beta_i$. As in the previous sections, the distance of a root from the unit circle $|\beta| = 1$ determines a decay scale. We define 
\begin{equation}
  \kappa_i
  =
  \left|\log\left|\beta_i\right|\right|,
  \label{eq:root_kappa}
  \end{equation}
  and the corresponding density decay length
  \begin{equation}
  \xi_i
  =
  \frac{1}{2\kappa_i}.
  \label{eq:root_lengths}
  \end{equation}
A root with $|\beta_i|<1$ decays away from the left boundary, while a root with
  $|\beta_i|>1$ gives the corresponding decay away from the right boundary. In
  both cases the decay rate is controlled by $\kappa_i$.

Because the coefficients of Eq.~\eqref{eq:mixed_root_polynomial} are real, non-real roots appear in complex-conjugate pairs. Such pairs have the same
  modulus and therefore the same $\kappa_i$ and $\xi_i$. We therefore group roots with the same value of $\kappa_i$ into physical decay channels, ordered by increasing decay exponent,
  \begin{equation}
  \kappa_{{\rm ch},1}
  \leq
  \kappa_{{\rm ch},2}
  \leq
  \cdots .
  \end{equation}
  The corresponding channel length is
  \begin{equation}
  \xi_{{\rm ch},a}
  =
  \frac{1}{2\kappa_{{\rm ch},a}} .
  \label{eq:mixed_channel_length}
  \end{equation}
Thus $\xi_{{\rm ch},1}$ is the longest distinct grouped-channel length. This grouping avoids treating the two members of one conjugate pair as two independent decay scales. 

We use the grouped-channel features to train random-forest regressors for $\ell_\star$. Here and in the following sections, numerical $\ell_\star$ denotes the value obtained by directly scanning over crop-lengths and selecting the smallest $\ell$ that satisfies the tolerance criterion, while predicted $\ell_\star$ denotes the output of the random-forest regressor. We consider both a global model, trained on all nonzero winding sectors together, and sector-resolved models, trained separately within each fixed winding sector. The sector-resolved models allow the relation between $\ell_\star$ and the leading decay length to vary between winding sectors.

\begin{figure}[t]
\centering
\includegraphics[width=0.95\linewidth]{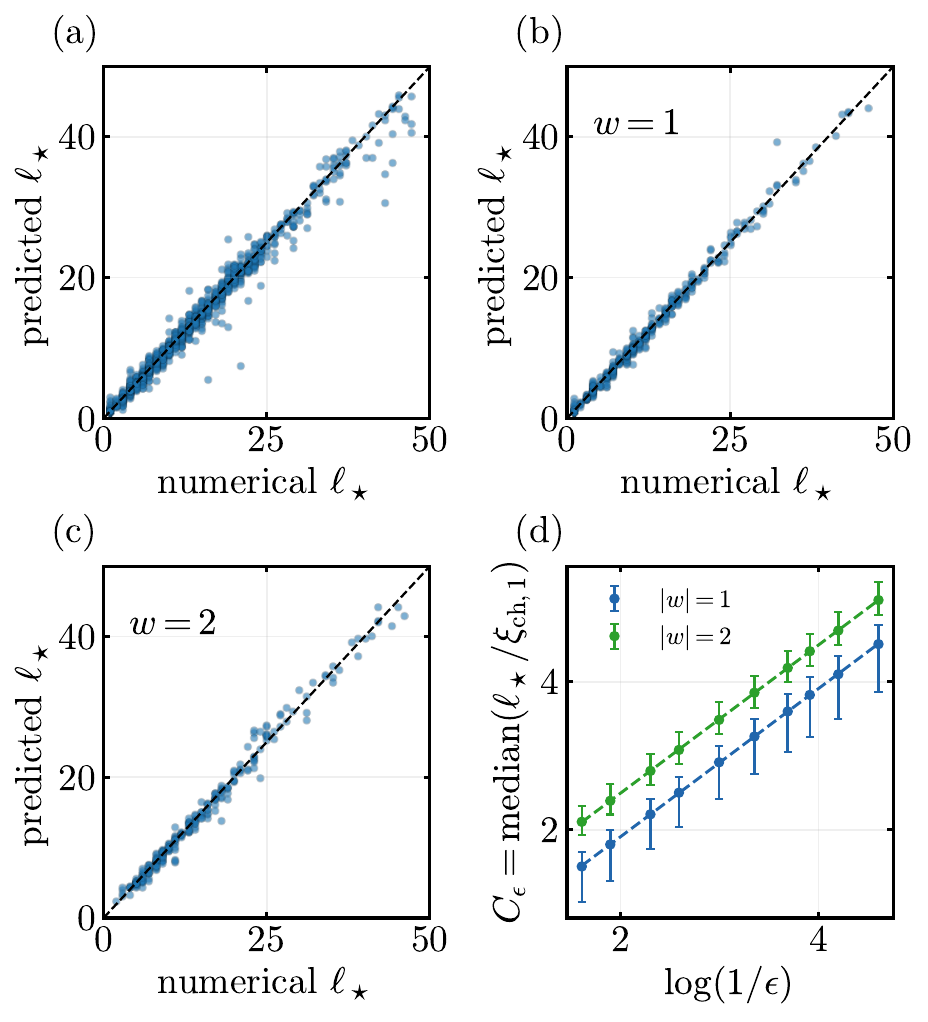}
\caption{
(a) Numerical $\ell_\star$ compared with the RF-predicted $\ell_\star$ from a random-forest model trained on all nonzero winding sectors. (b,c) Numerical $\ell_\star$ compared with the RF-predicted $\ell_\star$ from sector-resolved models for $w=1$ and $w=2$, respectively. The dashed line denotes perfect prediction.
(d) Tolerance dependence of $C_\epsilon={\rm median}(\ell_\star/\xi_{{\rm ch},1})$, where $\xi_{{\rm ch},1}$ is the length of the grouped root channel closest to the unit circle in the complex $\beta$ plane. Error bars indicate the interquartile range over sampled hopping parameters. Dashed lines are linear fits in $\log(1/\epsilon)$.
}
\label{fig:mixed_grouped_channels}
\end{figure}
The prediction results are shown in Fig.~\ref{fig:mixed_grouped_channels}. In all nonzero sectors, the dominant feature is the grouped channel closest to the unit circle, i.e. the channel with the smallest $\kappa_{{\rm ch},a}$. The sector-resolved regressions give $R^2=0.987,0.990,0.993,$ and $0.989$ for
  $w=-2,-1,1,$ and $2$, respectively. Thus, the same leading grouped channel controls all nonzero sectors, while the relation between this channel and $\ell_\star$ depends on the winding sector.

We also express the tolerance $\epsilon$ dependence in terms of the closest grouped-channel length. For each $\epsilon$, we define
\begin{equation}
C_\epsilon
=
{\rm median}
\left(
\frac{\ell_\star(\epsilon)}{\xi_{{\rm ch},1}}
\right),
\label{eq:mixed_ceps_grouped_definition}
\end{equation}
where the median is taken over the sampled hopping parameters in a fixed winding sector. The median is used because $\ell_\star$ is integer-valued and the mixed model has a broader distribution of crop-lengths than the pure hopping-range models.

As shown in Fig.~\ref{fig:mixed_grouped_channels}(d), $C_\epsilon$ grows approximately linearly with $\log(1/\epsilon)$. Thus the mixed model obeys the grouped-channel scaling form
\begin{equation}
\ell_\star(\epsilon)
\approx
C_\epsilon \xi_{{\rm ch},1},
\qquad
C_\epsilon
\simeq
a_{|w|}\log(1/\epsilon)+b_{|w|}.
\label{eq:mixed_grouped_lstar_scaling}
\end{equation}
The fitted coefficients are $(a_1,b_1)=(1.001,-0.096)$ for $|w|=1$ and $(a_2,b_2)=(1.001,0.493)$ for $|w|=2$. The sector dependence appears mainly in the fitted prefactor $C_\epsilon$, not in a different leading decay length. An integer crop estimate can then be obtained by taking the ceiling of the continuous value in Eq.~\eqref{eq:mixed_grouped_lstar_scaling}.

Thus, the mixed model differs from the pure $m$-hop models because the root equation contains several distinct decay channels. After grouping conjugate roots, however, the dominant scale is again simple: $\ell_\star$ is primarily controlled by the grouped root channel closest to the unit circle.

\section{Complex base-point convergence}
\label{sec:complex_EB_extension}

In Secs.~\ref{sec:nn_hn} and~\ref{sec:mhop_hn}, we solved the crop-length problem at the base point $E_B=0$. We now extend the same question to general $E_B\in\mathbb{C}$, both inside and outside the point-gapped spectral loop. For each $E_B$, the true value of the winding $w(E_B)$ is computed using Eq.~\eqref{eq:kspace_winding_general}.

Drawing from the insight that root-derived features provide enough information to predict $\ell_\star$ for the pure $m$-hop model at $E_B=0$ (see Sec.~\ref{sec:mhop_hn}), we construct the corresponding root features for general complex base point $E_B$ by solving  
\begin{equation} \label{eq:m-hop_EB_Ham}
    J_R\beta^{-m}+J_L\beta^m-E_B=0.
\end{equation}
Multiplying by $\beta^m$ and defining $z=\beta^m$ gives
  \begin{equation}
      J_L z^2-E_B z+J_R=0,
  \end{equation}
  with solutions
  \begin{equation}
      z_\pm(E_B)
      =
      \frac{E_B\pm\sqrt{E_B^2-4J_LJ_R}}{2J_L}.
      \label{eq:mhop_z_roots_general}
  \end{equation}
  Each solution $z_\pm$ generates $m$ roots of $\beta^m=z_\pm$, all with the same modulus. Thus the $2m$ roots are described by two radial decay exponents,
  \begin{equation}\label{eq:mhop_complex_kappa_pm}
      \kappa_\pm(E_B)
      =
      \frac{1}{m}\left|\log |z_\pm(E_B)|\right|.
  \end{equation}

The complex-$E_B$ data were generated at $J_R=1$ for
  $J_L=0.4,0.5,0.6,0.7,0.8$. The system sizes were $N=256,300,402,500$ for $m=1,2,3,4$, respectively. The target crop-length
  $\ell_\star$ is defined as the smallest crop-length for which $|w_{\rm PD}(E_B)-w(E_B)|<\epsilon$, with $\epsilon=0.10$.

For each fixed $m$, we use a leave-one-$J_L$-out validation protocol. The model is trained on four values of $J_L$ and tested on the remaining value (held-out value), so that the test data use a hopping strength not seen during training. This procedure is repeated for each of the five possible held-out values of $J_L$. The final scores are averaged over these five train-test runs (folds) and over $m=1,2,3,4$.

  Given the success of the simple scaling form at $E_B=0$ away from the transition point $|J_L|=|J_R|$ (see Eq.~\eqref{eq:mhop_lstar_xi}), we
  first tested the proportional baseline
  \begin{equation}
      \ell_\star \simeq C_\epsilon \xi_{\rm slow},
  \end{equation}
  where
  \begin{equation}
      \xi_{\rm slow}
      =
      \max_{\pm}
      \left[
      \frac{1}{2 \kappa_\pm(E_B)}
      \right].
  \end{equation}
  In each leave-one-$J_L$-out train-test split, the single coefficient $C_\epsilon$ was fitted using the four training values of $J_L$ and then tested on the held-out value. This captures the broad scaling away from the spectral boundary, but fails close to it. Averaged over all folds and all four
  values of $m$, this proportional fit gives $R^2 = 0.927$.

  We next applied the same validation protocol to a finite-size rational
  baseline,
  \begin{equation}
      \frac{\ell_\star}{N}
      =
      \frac{a}{N/\xi_{\rm slow}+c}
      + b .
  \end{equation}
  For each held-out fold, the parameters $(a,c,b)$ were fitted using the four
  training values of $J_L$ and then evaluated on the held-out value. This
  improves the average score to $R^2 = 0.984$, but errors remain near the spectral boundary.

\begin{figure}[t]
  \centering
  \includegraphics[width=0.95\columnwidth]{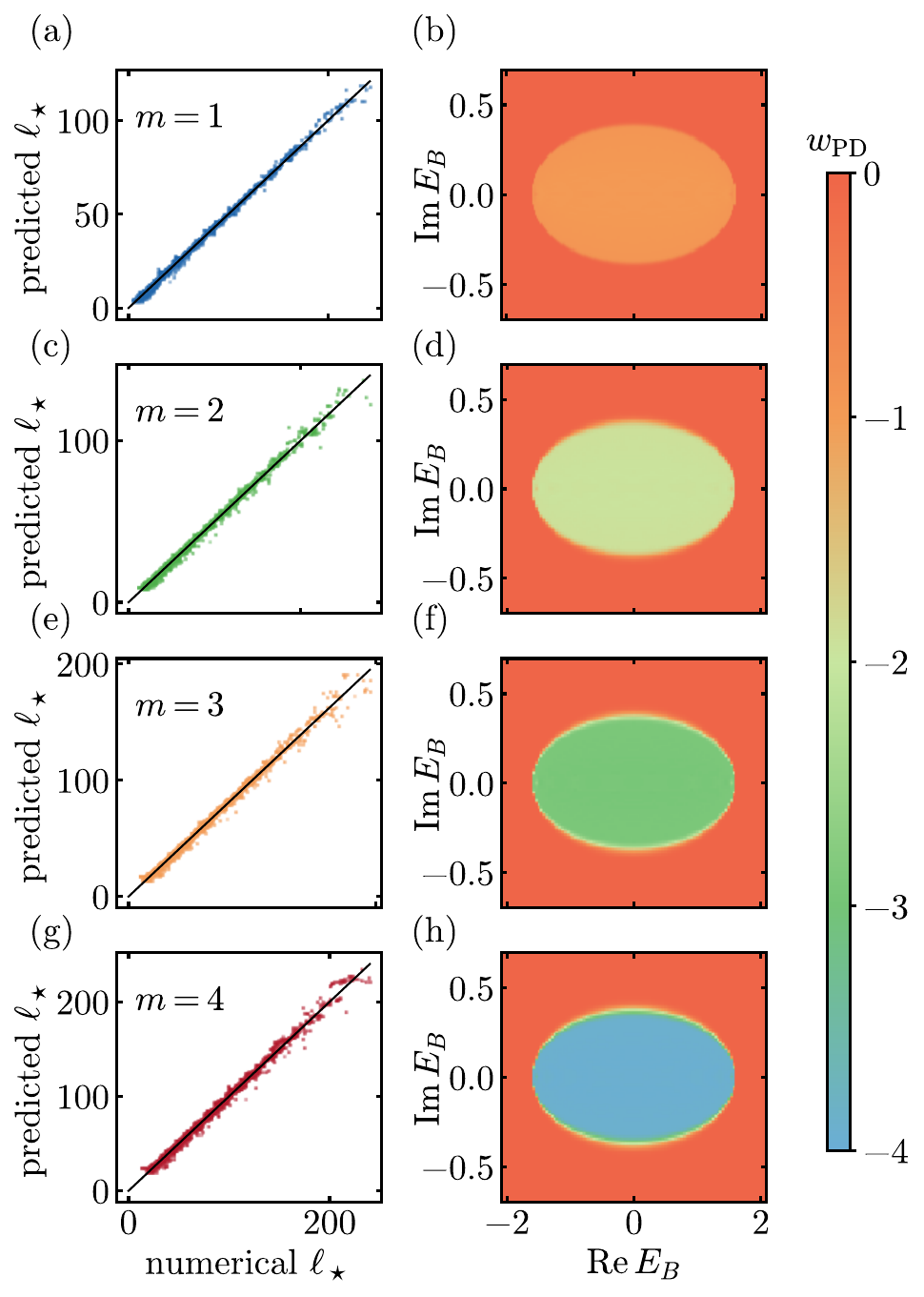}
  \caption{  Random-forest prediction of the crop-length $\ell_{\star}$ for $m$-hop 1D non-Hermitian chains (see Eq.~\ref{eq:m-hop_EB_Ham}). Rows correspond to $m=1,2,3,4$.
  Panels (a,c,e,g) compare the numerical $\ell_\star$ with the RF-prediction from grouped root-channel features; the black line indicates perfect
  agreement. Panels (b,d,f,h) show the corresponding reconstructed $w_{\rm PD}$ maps over the complex base-energy plane for the held-out
  value $J_L=0.6$.}
  \label{fig:pure_mhop_complex_EB}
  \end{figure}

  Finally, we used the same leave-one-$J_L$-out splits to train a
  random-forest regressor. In this case, the input features were the grouped
  root-channel quantities from the moduli and phases of the $\beta$ roots. Averaged over all four $m$ values and all held-out
  $J_L$ folds, the RF-predictor gives $R^2=0.993$.
  The resulting numerical-versus-predicted comparison and reconstructed phase maps are shown in Fig.~\ref{fig:pure_mhop_complex_EB}; the displayed
  maps use the representative held-out value $J_L=0.6$.
  
  Thus the roots of $H(\beta)-E_B$ provide effective features for predicting the crop-length across the complex-$E_B$ plane, while the random-forest captures near-boundary corrections beyond the simple one-length scaling form. 

The pure $m$-hop model has a constrained two-family root structure. We now study the mixed model introduced in Sec.~\ref{sec:mixed_model} and solve
  the crop-length problem at general complex $E_B\in\mathbb{C}$. For a
  complex base point $E_B$, the equation $H(\beta)-E_B=0$ becomes
  \begin{equation} \label{eq:2ndnnhop_EB_Ham}
  J_{L2}\beta^4
  +J_{L1}\beta^3
  -E_B\beta^2
  +J_{R1}\beta
  +J_{R2}
  =0 .
  \end{equation}
  This gives four roots, denoted $\beta_a$. We describe their radial positions by 
  \begin{equation}
      \rho_a=\log|\beta_a|,
  \end{equation}
  and the corresponding decay lengths by
  \begin{equation}
      \xi_a=\frac{1}{2|\rho_a|}.
  \end{equation}
  $\rho_a>0$ for roots outside the unit circle $|\beta_a| = 1$ and $\rho_a<0$ for roots inside it. Thus $\xi_a$ measures only the distance from the unit circle, while $\rho_a$ also keeps the inside/outside sign information.

To test the mixed hopping model, we selected eight representative hopping sets $(J_{L1}, J_{R1}, J_{L2}, J_{R2})$ whose Bloch Hamiltonians (see Eq.~\eqref{eq:mixed_bloch}) form non-trivial point-gapped loops with different shapes in the complex-energy plane. For each such Hamiltonian, we scan $E_B$ over the complex plane and compute $\ell_\star(E_B)$ at $N=256$ and $\epsilon=0.05$. Six loop geometries are used for training, and the remaining two are held out for testing.

\begin{figure}[t]
  \centering
  \includegraphics[width=0.95\columnwidth]{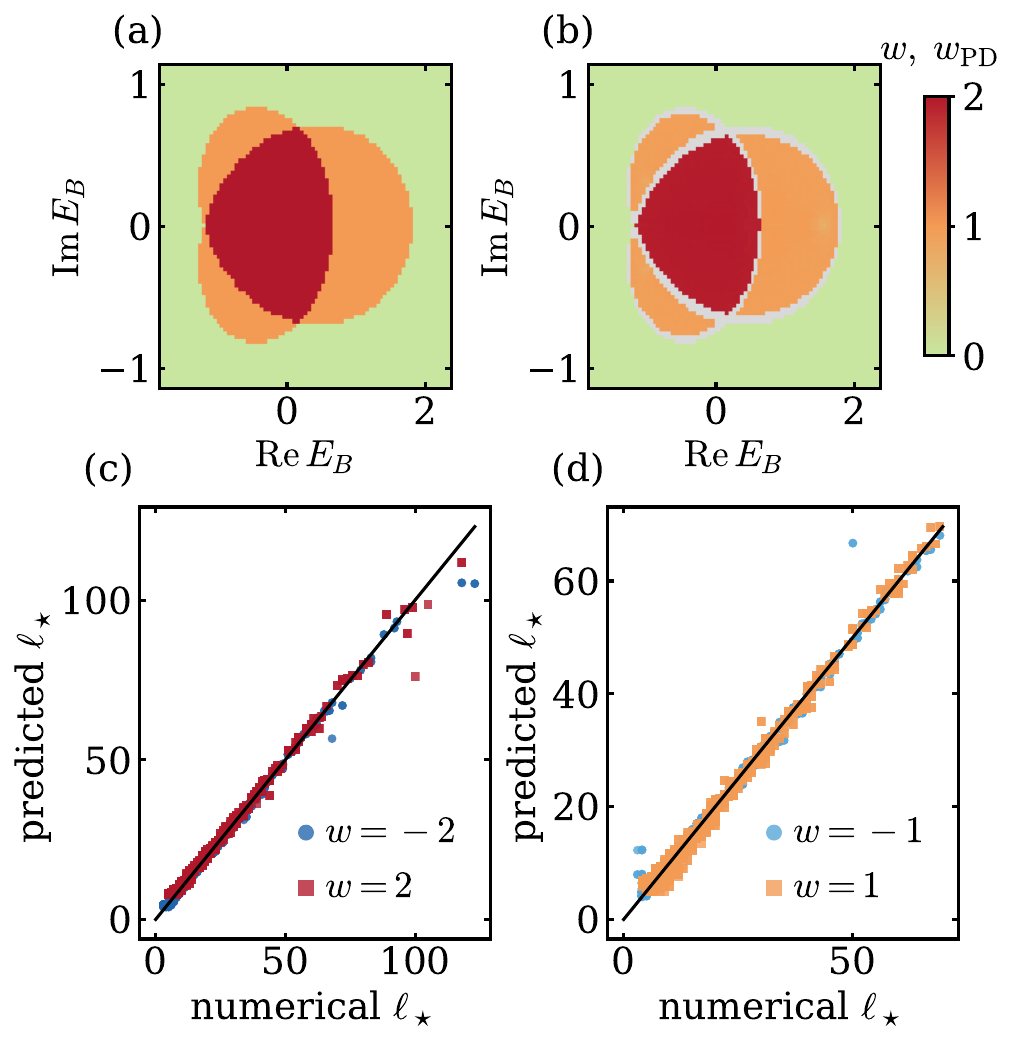}
  \caption{
  (a) Exact $k$-space winding $w(E_B)$ over the complex
  energy plane for one representative sample. (b) Reconstructed real-space invariant $w_{\rm PD}$ evaluated using the RF-predicted crop-length. Gray points denote base energies for which no valid crop-length exists at this $N$ and tolerance, meaning that the finite chain does not reach
  $\Delta (\ell)<\epsilon$ within the allowed crop window. (c,d) Numerical and RF-predicted crop-lengths in the $|w|=2$ and $|w| = 1$ sectors, respectively. The diagonal line indicates perfect prediction. 
  }
  \label{fig:mixed_complex_EB_clean}
  \end{figure}

We first test a one-length predictor for the mixed hopping model, using only the largest root-derived length scale,
  \begin{equation}
      \xi_{\rm ch1}=\max_a \frac{1}{2|\rho_a|},
  \end{equation}
  which corresponds to the root closest to the unit circle. This feature already
  predicts $\ell_{\star}$ well in the outer $|w|=2$ sectors, giving $R^2=0.992$ for
  $w=-2$ and $R^2=0.987$ for $w=+2$, but fails in the intermediate $w=\pm1$ sectors between the $w=0$ and $w=\pm2$ regions (see Fig.~\ref{fig:mixed_complex_EB_clean}). For
  these sectors, the one-length predictor gives only $R^2=0.179$ for $w=-1$ and $R^2=0.174$ for $w=+1$.

  We then train the RF on the full signed radial root vector
  \begin{equation}
      (\rho_1,\rho_2,\rho_3,\rho_4),
  \end{equation}
  with the roots ordered by increasing $|\rho_a|$. This retains both the distance of each root from the unit circle and whether it lies inside or
  outside it. With this full root vector, the held-out predictions are accurate in all nonzero winding
  sectors: $R^2=0.995,0.994,0.992$, and $0.992$ for $w=-2,-1,+1,+2$, respectively.

Full $k$-space winding maps in the complex energy plane can be reconstructed by $w_{\rm PD}$ calculated for a finite chain using the RF-predicted $\ell_{\star}$. This is shown in Figure~\ref{fig:mixed_complex_EB_clean} (a-b) for one held-out mixed hopping Hamiltonian at $N=256$ and $\epsilon=0.05$. Moreover, the RF-predicted and numerical $\ell_{\star}$ match well on unseen loop geometries (see Figure~\ref{fig:mixed_complex_EB_clean} (c-d)).

\begin{figure}[t]
    \centering
    \includegraphics[width=0.95\columnwidth]{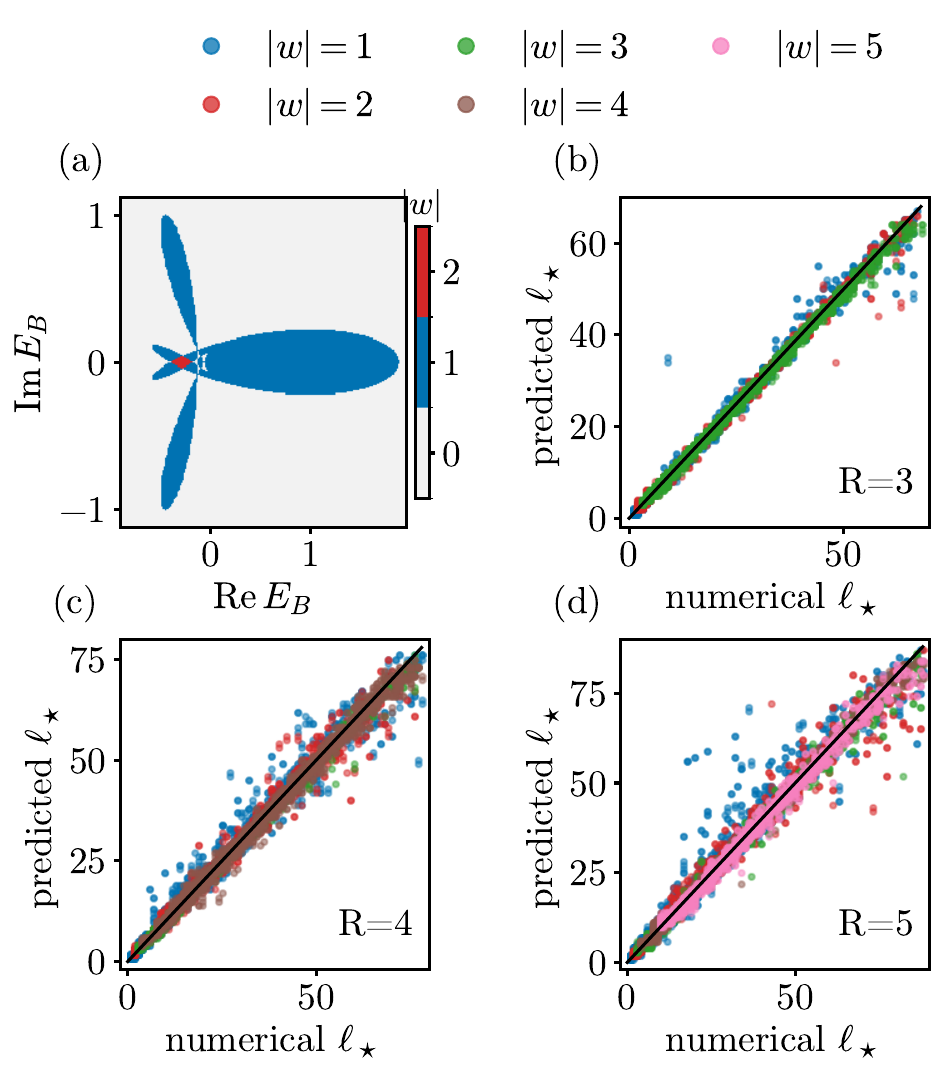}
    \caption{
     (a) Winding map $|w(E_B)|$ for $R=4$ Bloch Hamiltonian with $(J_{R,1},J_{R,2},J_{R,3},J_{R,4})=(0.162,0.442,0.078,0.338)$ and
  $(J_{L,1},J_{L,2},J_{L,3},J_{L,4})=(0.403,0.039,0.357,0.052)$.
 (b-d) Numerical crop-length $\ell_\star$, obtained by direct scans of the real-space invariant over $\ell$, compared with the RF-predicted value for held-out Hamiltonians with hopping ranges $R=3,4,5$, respectively. The diagonal line indicates perfect prediction.}
    \label{fig:finite_range_extension}
  \end{figure}
  
Finally, we extend this analysis from the first-/second-neighbor mixed model to more general mixed-hopping models with maximum hopping range $R$. The Bloch Hamiltonian for these models takes the form 
\begin{equation}
  H_R(k)
  =
  \sum_{r=1}^{R}
  \left(
  J_{Rr}e^{-irk}
  +
  J_{Lr}e^{irk}
  \right),
  \label{eq:finite_range_bloch}
  \end{equation}
  or, equivalently,
  \begin{equation}
  H_R(\beta)
  =
  \sum_{r=1}^{R}
  \left(
  J_{Rr}\beta^{-r}
  +
  J_{Lr}\beta^r
  \right).
  \label{eq:finite_range_beta}
  \end{equation}
The mixed hopping model in Eq.~\eqref{eq:mixed_bloch} is the special case $R = 2$. For each base point $E_B$, the equation $H_R(\beta)-E_B=0$ becomes a polynomial of degree $2R$ after multiplying by $\beta^R$. We compute all $2R$ roots and order them by increasing distance from the unit circle. 

The RF input contains the hopping amplitudes, the base point $E_B$, the
  clean winding sector, and the ordered root data 
  \begin{equation}
      \rho_i=\log |\beta_i|
  \end{equation}
  together with the corresponding root phases.

For each $R$, we generate independent clean hopping Hamiltonians and compute
  $\ell_\star(E_B)$ on a complex-$E_B$ grid. We train on one set of
  Hamiltonians and test on a separate held-out set. The training sets contain
  260, 320, and 220 Hamiltonians, and the test sets contain 24, 36, and 44
  Hamiltonians for $R=3,4,5$, respectively.

Figure~\ref{fig:finite_range_extension} compares the RF-predicted $\ell_\star$ with the numerical $\ell_\star$ obtained from direct crop-length scans, for valid points in nonzero topological sectors. The points follow the diagonal trend for all three hopping ranges, with the largest scatter appearing for $R=5$. This is expected because larger hopping range gives a longer root feature vector and more diverse loop geometries. Even in these cases, the predicted crop-lengths remain close to the numerical values for most valid points, with $R^2=0.996,0.994,0.991$ for $R=3,4,5$, respectively. Thus the same
  signed full-root representation remains effective for finite-range clean
  hopping models beyond the nearest-neighbor and first-/second-neighbor cases.

\begin{figure}[t]
      \centering
      \includegraphics[width= 0.8\columnwidth]{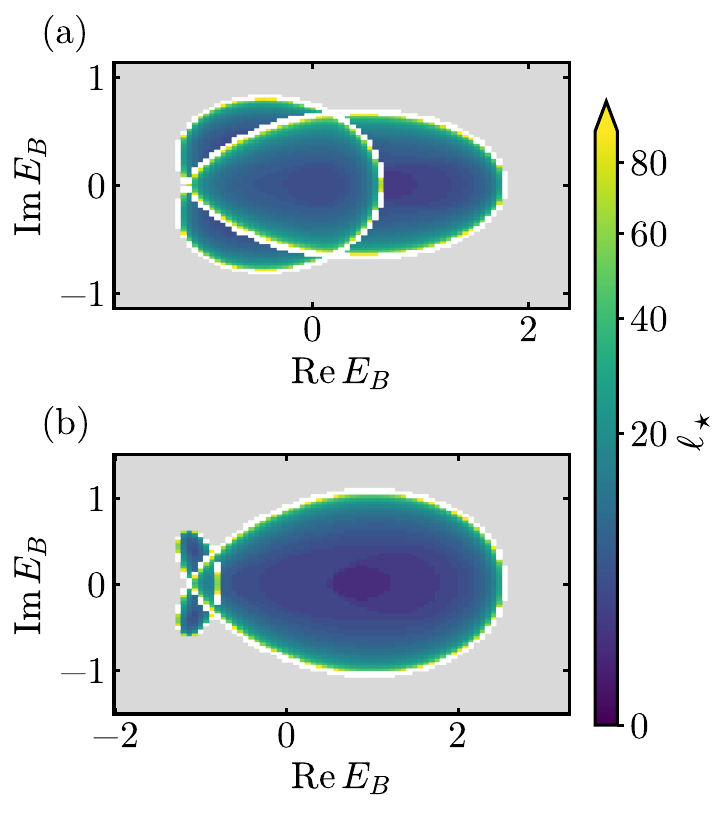}
      \caption{
      Numerical crop-length $\ell_{\star}(E_B)$ for two mixed-hopping Hamiltonians with (a) $(J_{R1},J_{L1},J_{R2},J_{L2})=(0.328,0.238,0.234,1)$ and (b) $(0.668,1,0.022,0.877)$, respectively. Here $N=256$ and $\epsilon=0.1$.
      Gray regions denote the trivial $w(E_B)=0$ sector, while white pixels near the loop boundaries correspond to grid points for which no valid crop-length exists within the allowed crop window.
      }
      \label{fig:lstar_EB_maps}
  \end{figure}

 To visualize what the RF regression is learning in the complex-energy extension, we also plot the numerical crop-length directly in the $E_B$ plane for two representative mixed first-/second-neighbor loops (see Fig.~\ref{fig:lstar_EB_maps}). Away from the point-gap boundary, only a modest crop is required, whereas $\ell_\star$ grows strongly as $E_B$ approaches the boundary of a nontrivial sector. This is consistent with the idea that the relevant decay length increases and a larger boundary crop is needed before $w_{\rm PD}$ reaches its quantized value.

 \section{Robustness against disorder} \label{sec:disorder_robustness}

In this section, we test the robustness of the crop-length predictor for the clean Hatano-Nelson model ($m = 1$) against hopping disorder at fixed system size $N=256$. For each base point $E_B$, the reference integer $w(E_B)$ is computed from the clean Bloch Hamiltonian (see Eq.~\eqref{eq:hn_nn_bloch}), and $\ell_\star^{\rm clean}(E_B)$ is predicted using the root-based RF-predictor. We then add disorder to the real-space hopping matrix and evaluate $w_{\rm PD}$ using $\ell_\star^{\rm clean}(E_B)$. 

\begin{table}[t]
  \centering
  \caption{
  Robustness of the clean crop-length predictor at $r_W=0.3$. Entries are success
  fractions for $\Delta W<0.1$ at $N=256$.}
  \label{tab:disorder_transfer_jl}
  \begin{tabular}{c c c c}
  \hline\hline
  $J_L$ & all gapped & interior & boundary \\
  \hline
  $0.5$ & $0.904$ & $0.965$ & $0.553$ \\
  $0.6$ & $0.906$ & $0.970$ & $0.512$ \\
  $0.7$ & $0.938$ & $0.987$ & $0.634$ \\
  \hline\hline
  \end{tabular}
  \end{table}

The disordered open-boundary Hamiltonian is
\begin{equation}
  H_{\rm dis}
  =
  \sum_n
  \left[
  J_R^n c_{n+1}^\dagger c_n
  +
  J_L^n c_n^\dagger c_{n+1}
  \right],
  \end{equation}
  with
  \begin{equation}
  J_R^n=J_R+W_R\omega_R^n,
  \qquad
  J_L^n=J_L+W_L\omega_L^n .
  \end{equation}
  Here $\omega_R^n$ and $\omega_L^n$ are independent random variables
  uniformly distributed in $[-1/2,1/2]$. We parameterize the disorder strength
  by
  \begin{equation}
      W_R=W_L=2r_W |J_R-J_L| .
  \end{equation}
  With this convention, the random correction to each hopping lies in
  $[-r_W|J_R-J_L|,\,r_W|J_R-J_L|]$. Thus $r_W$ is the maximum disorder
  amplitude measured in units of the clean hopping asymmetry $|J_R-J_L|$.

 \begin{figure}[t]
      \centering
      \includegraphics[width=0.95\columnwidth]{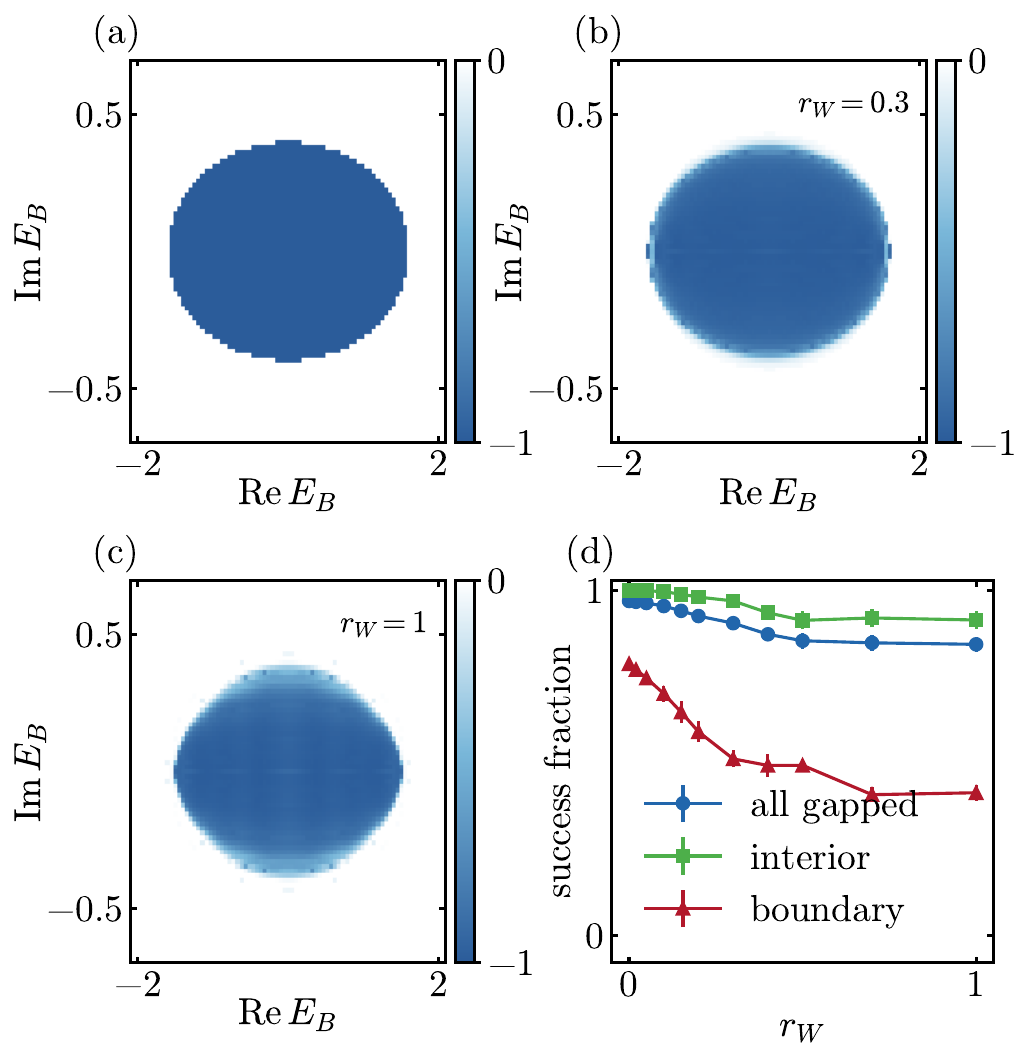}
      \caption{
      (a) Exact k-space winding $w(E_B)$ map for a clean Hatano-Nelson chain at $J_R=1$, and $J_L=0.6$. (b,c)
      $w_{\rm PD}^{\rm dis}(E_B;\ell_\star^{\rm clean})$ map for representative disorder realizations at $r_W=0.3$ and $r_W=1.0$, evaluated using the
      corresponding clean crop-length $\ell_\star^{\rm clean}$. The chain length is $N=256$. (d) Mean fraction of $E_B$ points satisfying $\Delta W<0.1$ as a function of $r_W$. The three curves show all clean-gapped points, phase-region interior points, and points near phase boundaries.
      Error bars denote standard errors of the mean across ten disorder
      realizations.}
      \label{fig:disorder_transfer}
  \end{figure}

  For each disorder realization, we compute $w_{\rm PD}^{\rm dis}\left(E_B;\ell_\star^{\rm clean}\right)$
  from the polar decomposition of $H_{\rm dis}-E_B$. We compare $w_{\rm PD}^{\rm dis}$ with the clean reference winding through
  \begin{equation}
      \Delta W(E_B)
      =
      \left|
      w_{\rm PD}^{\rm dis}
      \left(E_B;\ell_\star^{\rm clean}\right)
      -
      w(E_B)
      \right| .
  \end{equation}
  A base point is counted as successful when $\Delta W(E_B)< \epsilon = 0.1$.

Figure~\ref{fig:disorder_transfer} shows that the clean crop-length predictor
  remains useful under moderate hopping disorder. For $r_W=0.3$,
  $w_{\rm PD}^{\rm dis}(E_B;\ell_\star^{\rm clean})$ agrees with
  $w_{\rm clean}(E_B)$ over most of the clean phase-region interior. The largest
  deviations occur near the clean phase boundary, where the reference winding is most sensitive to changes in $E_B$. At $r_W=1.0$, the reconstruction degrades, but the contrast between interior and near-boundary points
  remains visible.

  To quantify this effect, we split the clean-gapped grid points into
  near-boundary and interior subsets. A point is classified as near-boundary if
  any nearest or second-nearest neighbor on the complex-$E_B$ grid has a
  different value of $w$; the remaining clean-gapped points are
  classified as interior. For each $r_W$, we compute the fraction of points
  satisfying $\Delta W(E_B)<0.1$ for ten independent disorder realizations and
  plot the mean and standard error in Fig.~\ref{fig:disorder_transfer}(d). The
  interior points remain more stable than the near-boundary points as the
  disorder strength is increased.

  We repeated the same test for three clean nearest-neighbor Hatano-Nelson
  loops with $J_R=1$ and different values of $J_L$, at fixed $r_W=0.3$. The
  results are summarized in Table~\ref{tab:disorder_transfer_jl}. In all three
  cases, the success fraction for the interior points remains above $0.96$,
  while the success fraction close to the phase boundary is lower. This is
  consistent with the clean finite-size picture: close to the phase boundary,
  the point gap is small and the reconstruction is more sensitive to disorder.

\section{Conclusions}
\label{sec:conclusion}

 In finite non-Hermitian chains with the non-Hermitian skin effect, the polar-decomposition real-space invariant depends sensitively on the choice of a cutoff parameter. This parameter, which we termed the crop-length, controls the convergence of the invariant to the corresponding momentum-space winding number. Especially close to topological phase transitions, where finite-size effects are strongest, choosing this parameter becomes very important.

In this study, we showed that this convergence parameter is controlled by decay lengths associated with the non-Hermitian skin effect. We defined $\ell_\star$ as the smallest crop-length for which the real-space topological invariant agrees with the momentum-space winding number within a fixed tolerance $\epsilon$. We used random-forest classifiers to determine whether this tolerance can be reached within the allowed crop range, and random-forest regressors to predict $\ell_\star$ when it can. This supervised learning method was tested across finite non-Hermitian chains with different hopping ranges, winding sectors, complex base energies, and disorder
  strengths.

 For nearest-neighbor Hatano-Nelson chains, the Bloch spectrum forms an ellipse in the complex-energy plane. Taking $E_B = 0$ as the base point, we showed that the required crop-length is controlled by the skin-effect localization length $\xi=1/|\log|J_R/J_L||$, where $J_R$ and $J_L$ are non-reciprocal hoppings.  Away from the transition, $\ell_\star$ is approximately proportional to $\xi$, so the convergence of the real-space invariant is set by the same decay scale that controls the non-Hermitian skin effect. For $\epsilon=0.05$, this gives the practical requirement
  $N\gtrsim 10\xi$ for the existence of a valid crop, where $N$ is the size of the chain. For the pure $m$th-neighbor hopping model, the spectrum has the same ellipse-like geometry but winds $m$ times around the base point. At $E_B = 0$, the same crop-length picture holds after replacing $\xi$ by $\xi_m=m/|\log|J_R/J_L||$. In both cases, the random-forest models identify these decay lengths as the main predictors of $\ell_{\star}$.

For mixed first- and second-neighbor hopping models, the spectral loops are no longer simple ellipses. They can become strongly non-elliptic with deformed or multi-lobed shapes and nested winding sectors. In this case, the crop-length
  problem involves several decay channels rather than a single localization length. At base point $E_B=0$, these channels
  are obtained from the roots $\beta$ of the characteristic equation
  $H(\beta)-E_B=0$: grouping complex-conjugate roots gives the relevant physical channels, and the slowest decaying channel controls the dominant trend.
  
  For general complex base points, the predictor needs the signed radial root information in order to distinguish the nested topological winding sectors. This is visible in the nontrivial phase diagrams with strongly non-elliptic spectral loops, where a single-length predictor is no longer sufficient. The same signed full-root representation also works for finite-range clean hopping
  models with hopping ranges $R=3,4,5$.

  We also tested the clean nearest-neighbor crop predictor under hopping disorder. The reference value was kept fixed as the momentum-space winding number, and the disordered real-space calculation used the crop-length predicted from the clean system. The predictor remains stable in the interior of the clean phase regions under moderate disorder, while deviations
  are concentrated near phase boundaries where the point gap is small and the finite-size sensitivity is largest.
  
Our results provide a practical way to make the topological invariant reliable for finite non-Hermitian chains with distinct topological phases. Moreover, the learned relations between the crop-length parameter, tolerance, and the decay scales can be used to estimate effective skin-effect decay lengths in different non-Hermitian Hamiltonians, and also in experimentally obtained non-Hermitian matrices.

\section{Data availability}
The data and code that support the findings of this study are available on Zenodo \cite{Zenodo2026}. 

\section{Acknowledgments}

We acknowledge the support of the German Research Foundation (DFG) through the Collaborative Research Center ToCoTronics, Project-ID 258499086 - SFB 1170, as well
as Germany’s Excellence Strategy through the W\"{u}rzburg-Dresden Cluster of Excellence on Complexity and Topology in Quantum Matter - ctd.qmat (EXC 2147, Project-ID
390858490).

\bibliography{main}

\appendix

\section{Finite-size branches used in the train-test split}
\label{app:finite_size_branches}

In Secs.~\ref{sec:nn_hn} and~\ref{sec:mhop_hn}, the train-test split is performed within fixed finite-size branches. This section explains the numerical motivation for this choice. The branch structure is not used as an additional scaling law for $\ell_\star$. It is only used to avoid mixing weak residue-class oscillations in the supervised-learning procedure.

For a fixed crop-length $\ell$, we define 
\begin{equation}
\Delta_\ell(N)=|w_{\rm PD}(\ell,N)-w|,
\end{equation}
where $w$ is the exact clean momentum-space winding. This fixed-crop quantity is different from the crop-length problem in the main text: here $\ell$ is fixed and $N$ is varied, whereas in the main analysis $\ell$ is varied to find the minimum value $\ell_\star$. 

\begin{figure}[t]
\centering
\includegraphics[width=0.95\linewidth]{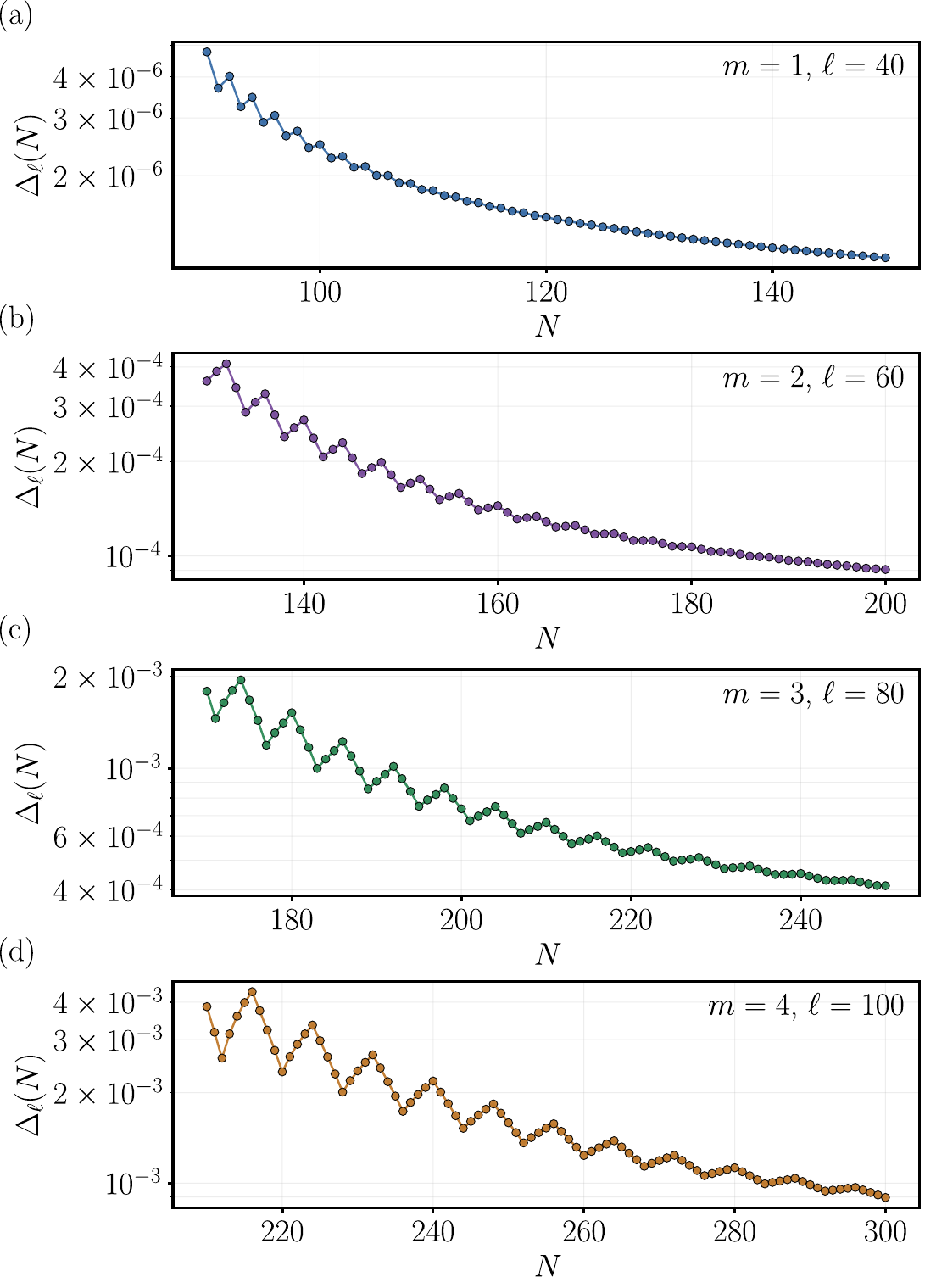}
\caption{
Fixed-crop finite-size diagnostic for pure $m$-th-neighbor HN chains. 
The plotted quantity is $\Delta_\ell(N)=|w_{\rm PD}(\ell,N)-w|$ at fixed crop-length $\ell$. 
Panels (a-d) show $m=1,2,3,4$, respectively. 
For each $m$, the envelope decreases with increasing $N$, while the error shows a visible oscillation. 
The observed period is $2m$, namely $2,4,6,$ and $8$ for the four panels. 
This diagnostic motivates the $N\bmod2m$ branch-aware train-test split used in the ML analysis.
}
\label{fig:appendix_branch_diagnostic}
\end{figure}

Fig.~\ref{fig:appendix_branch_diagnostic} shows $\Delta_\ell(N)$ for $m = 1,2,3,4$. In all cases, the envelope decreases with increasing $N$, but the error is not strictly monotonic. Instead, points with different values of $N \bmod 2m$ form weak oscillatory branches. The oscillation amplitude decreases
  with increasing $N$, so this is a finite-size effect and does not change the
  main decay-length scale controlling $\ell_\star$.

For the nearest-neighbor chain, the branch label is $N\bmod 2$. For pure
$m$-th-hop chains, we use $N\bmod 2m$. Within each branch, the train-test split is performed separately. The ML scores reported in the main text are then obtained by averaging over the branches, with the spread reported as a standard deviation.

\end{document}